# *The User-first Approach to AI Ethics: Preferences for Ethical Principles in AI Systems across Cultures and Contexts*

Benjamin J. Carroll, University of Technology Sydney, Australia, benjamin.j.carroll@uts.edu.au (corresponding author)

Jianlong Zhou, University of Technology Sydney, Australia, jianlong.zhou@uts.edu.au

Paul F. Burke, University of Technology Sydney, Australia, paul.burke@uts.edu.au

Sabine Ammon, Technical University Berlin, Germany, ammon@tu-berlin.de

**Abstract**

As AI systems increasingly permeate everyday life, designers and developers face mounting pressure to balance innovation with ethical design choices. To date, the operationalisation of AI ethics has predominantly depended on frameworks that prescribe which ethical principles should be embedded within AI systems. However, the extent to which users value these principles remains largely unexplored in the existing literature. In a discrete choice experiment conducted in four countries, we quantify user preferences for 11 ethical principles. Our findings indicate that, while users generally prioritise privacy, justice & fairness, and transparency, their preferences exhibit significant variation based on culture and application context. Latent class analysis further revealed four distinct user cohorts, the largest of which is ethically disengaged and defers to regulatory oversight. Our findings offer (1) empirical evidence of uneven user prioritisation of AI ethics principles, (2) actionable guidance for operationalising ethics tailored to culture and context, (3) support for the development of robust regulatory mechanisms, and (4) a foundation for advancing a user-centred approach to AI ethics, motivated independently from abstract moral theory.

---

## Introduction

In recent years, as capabilities have improved, artificial intelligence (AI) has become increasingly ubiquitous. Presently, in the United States (US) alone, hundreds of AI-enabled medical devices are approved for market each year, and self-driving cars provide hundreds-of-thousands of autonomous trips each week (Maslej et al. 2025). Recent global surveys reveal that more than half of respondents report using AI on a regular basis, including in their work (Gillespie et al. 2025), and business leadership report that AI is their top priority (KPMG 2024 CEO Outlook 2024). Indeed, AI is becoming embedded into virtually every aspect of modern society (Zhou and Chen 2023) , and most people expect AI will significantly impact their daily lives in the coming years (Maslej et al. 2025).

AI systems have the potential to have positive impact on people's lives and society. They can make time-consuming tasks easier (Benbya et al. 2024), provide cognitive support during difficult decisions (Carroll et al. 2025; Sadeghian and Otarkhani 2024), improve student learning outcomes (Wang and Fan 2025), assist with medical diagnosis (Salinas et al. 2024), and help with scientific writing (Salvagno et al. 2023). Furthermore, AI is expected to be broadly positive in the accomplishment of the United Nations (UN) Sustainable Development Goals (SDGs) (Satornino et al. 2024; Vinuesa et al. 2020).

Yet, despite this, there is evidence of general unease regarding AI systems. Popular chatbots such as ChatGPT have given people first-hand experience with AI errors such as hallucinations, which fuel mistrust (Rapp et al. 2025; Zhou, H. Müller, et al. 2024) and lower perceptions of AI capability (Lee et al. 2024). People report a sense of algorithm aversion towards AI-driven problem solving, even when AI systems perform better than human counterparts (Castelo et al. 2019; Dietvorst et al. 2015; Jussupow et al. 2020). At the societal level, AI assessments tend to have a negativity bias which focuses on AI failings (Singh Chauhan 2024). Afterall, AI systems have been found to discriminate against gender (Dastin 2022; Obermeyer et al. 2019; Prates et al. 2019), age (Kotwal and Marcel 2025), cause social harms (Michael 2024; Rinta-Kahila et al. 2024), and endanger human safety (Doyle 2023; Hong et al. 2021; Koopman 2024). While individual attitudes towards AI are nuanced (see arguments for algorithm appreciation, Logg et al. (2019), the moderating factor of expertise (Hou and Jung 2021), and a priori beliefs in AI strengths (Jago and Laurin 2022)), there is nevertheless a high level of AI skepticism that persists across cultures (Gillespie et al. 2025) and applications (Hengstler et al. 2016), which has increased in recent years (Maslej et al. 2025). This presents developers and designers of AI systems with a major challenge when designing AI systems for which user adoption is an objective.

In response to these concerns, AI ethics has emerged as an indispensable part of AI development (Kazim and Koshiyama 2021). Ethics is a branch of philosophy which considers what is morally right and wrong, or good and bad (Dewey and Tufts 1909). Applying ethics to AI systems requires consideration of the unique aspects of the technology such as scale, automation, and opacity, and how they impact individuals and society (V. C. Müller 2025). With this increasing presence of AI systems in our daily lives, AI ethics has become a global focus of academic researchers and policy makers (Mittelstadt 2019). This has resulted in the global production of hundreds of AI ethics frameworks (Fjeld et al. 2020). These documents, created by governments, non-profit groups, and corporations, identify which ethical principles[1] should be implemented in the development of AI systems. However, despite these efforts, there is limited evidence to suggest these documents have meaningfully contributed to changes in how AI systems are designed and developed (Hagendorff 2020). This problem of operationalisation[2] has persisted for some time, and can be thought of as a "lack of follow-through" (Zhou and Chen 2023, p. 2697).

This problem of AI ethics operationalisation can perhaps be understood in the context of the shift from AI being a research pursuit led by universities, to one dominated by corporations (De Sousa 2025; Owens 2024). Operating in competitive environments where duty to shareholders is primary, it should be no surprise that commercial interests take primacy over ethical concerns (Mittelstadt 2019). AI ethics conflict with the profit motive because they are technically challenging to implement, require making complex trade-offs, and involve financial outlay (Lu et al. 2022).

1 We use principles, values, and attributes interchangeably in this paper to refer to ethical concepts as they relate to AI systems.

2 We use operationalisation in this paper to refer to both the process of translating abstract ethical concepts to concrete design principles, as well as the process of actually implementing them in AI systems.

Indeed, AI developers report pressure to focus on performance and innovation over ethical concerns (Browne et al. 2024; Griffin et al. 2024; Ibáñez and Olmeda 2022; Morley et al. 2023). The prioritisation of corporate profits over ethical concerns is hardly a new phenomenon (M. Friedman 1970), however, in the case of AI ethics it may be short sighted. Embedding ethical values into the design and development of AI systems has been found to drive consumer trust and satisfaction (Capgemini 2022; Gillespie et al. 2025; Glikson and Woolley 2020; Kaplan et al. 2023), which in turn meaningfully and positively influences the intent to use such systems (Shao et al. 2024). If an organisation's profit-driven goal with regards to AI is derived from its use and adoption, then AI ethics must be part of the plan.

We believe that the approach to AI ethics operationalisation to date has struggled because it has not recognised that modern AI systems are for-profit services focused on consumer end-users. Absent binding regulations or a legal framework, this top-down approach, which utilises prescriptive frameworks for AI ethics, is less relevant for businesses. We believe a solution to this problem lies in a bottom-up approach that starts with user preferences, and how AI ethics can be utilised by designers to foster user intent to use, adoption, and satisfaction. This provides a more robust basis for businesses to incorporate ethical design principles as it is based on business goals, rather than abstract moral theory.

The first part of this approach requires understanding which of the principles of AI ethics are important to consumers, and to what extent. This provides the basic information that will allow AI designers and developers to assess which ethical design principles should be prioritised and which trade-offs must be made. Hence, the following research question is proposed:

**RQ1**: How do AI ethical principles meaningfully differ in user preference?

In keeping with the consumer perspective, we anticipate that consumer preferences will be heterogeneous across consumer contexts, as suggested by Zhou and Chen (2023), and thereby propose the following for examination:

**RQ2**: How do these preferences change depending on context?

Additionally, we anticipate that ethical preferences will differ across cultures, as there is evidence that approaches to AI ethics differ between countries (Corrêa et al. 2023; Hongladarom and Bandasak 2024). We ask the following research question:

**RQ3**: How do these preferences change depending on culture?

Finally, we must consider the impact of general AI attitudes and AI literacy, as individual factors have been shown to be important in how users form attitudes towards AI systems (for instance, see Gursoy et al. (2019) and Kaplan et al. (2023)). Hence, we include the following two research questions for investigation:

**RQ4**: How do these preferences change depending on the user's general AI attitudes?

**RQ5**: How do these preferences change depending on the user's general AI literacy?

To address these questions, we conducted a discrete choice experiment (DCE) with 829 participants from four countries (United States, China, Germany, and Australia) to assess the relative preferences for each of the 11 AI ethical principles common to AI ethics frameworks globally. The DCE methodology is a robust stated choice methodology originally developed in econometrics and marketing science, and now widely used across a variety of disciplines to assess preferences for individual attributes of product offerings (J. J. Louviere et al. 2000). We conducted the experiment across three consumer contexts (medical, financial, educational) which reflect nascent AI applications and conducted analysis utilising Apollo choice modelling software (Hess and Palma 2019).

Our findings offer:

1) Empirical evidence of AI ethics preference heterogeneity at the individual level participants. We show that, at the aggregate level, the ethical principles of privacy, justice & fairness, and transparency were most favored, while sustainability and solidarity are relatively deprioritized. This suggests consumers would value the embedding of some ethical principles in AI systems more than other.
2) Actionable guidance for operationalising ethics tailored to country and context. We show that preferences for ethical attributes in AI systems differ meaningfully between application context and country, supporting the need for a context-specific strategy for implementing AI ethics.
3) Support for the development of regulatory mechanisms. Our analysis reveals four distinct ethical cohorts of users, the largest of which we label as ethically disengaged. This group, representing 47% of participants, displayed no specific preference for ethical principles other than responsibility. This suggests that the development of meaningful oversight mechanisms (see Cheong (2024), Dignum (2020), Kroll (2020), and Lechterman (2022)) may provide an appropriate alternative approach to AI ethics operationalisation.
4) A foundation for a user-first approach to AI ethics which is motivated independently of abstract moral theory. We identify that operationalising AI ethics in comtemporary systems require a motivation consistent with commercial needs of designers and developers. Our research shows that consumers have clear ethical preferences. This can be utilised by designers and developers in order to improve usage intent, adoption, and satisfaction.

Overall, our research supports the importance of cultural and context-specific approach to AI ethics, and suggest that globally standardised ethical frameworks may fail to resonate uniformly across user populations. Our research provides a clear contribution to literature by providing empirical evidence of heterogeneity in user preferences. Our findings support ongoing discussions in human-computer interaction (HCI) and human-AI interaction (HAII) literature, and provide practical guidance to practitioners on how to approach the challenges associated with AI ethics operationalisation.

## Theoretical background

### *Ethics & the design of AI systems*

Ethical philosophy dates back at least as far as ancient Greece, and concerns what is morally right and wrong, or good and bad (Dewey and Tufts 1909). As a body of knowledge, it helps people decide how they and others should behave (Dubber et al. 2020). The incorporation of ethics into human-computer interaction (HCI) literature is well established. Investigation and conversation about bias and fairness in system design, for instance, dates back at least 40 years. More recently, the volume of papers mentioning ’ethics’ or ’human values’ has far outpaced general literature growth rates (Shilton 2018). This focus on ethics and human values has led to established paradigms such as value-sensitive design (VSD) (B. Friedman et al. 2013), the establishment of conferences such as ACM FAccT (Fairness, Accountability, and Transparency). At a higher level, there is the notion that the next ’wave’ of HCI research will be one where ethics and values are central and at the forefront (Ashby et al. 2019; Lopes 2022). Many of these ideas apply to efforts to establish design guidelines for AI systems (Amershi et al. 2019).

Ethical design, then, is tightly interwoven with HCI research, and principles of AI system design. However, this is not merely an exercise in morality; research suggests that the perception of ethical design carries with it tangible, positive impacts on user behaviour. Applying the Technology Adoption Model (TAM) (Davis et al. 1989), Shao et al. (2024) found that the perception of ethical design improves both perceived ease of use, and perceived usefulness of the AI system, which positively impacts usage intentions. Aysolmaz et al. (2023) found a similar effect, showing that the perception of ethical principles of fairness, accountability, and privacy each had a positive effect on user trust and perceived usefulness of an ADM system, which in turn was positively associated with adoption intention. Shin (2020) looked at decisions made by personalised AI systems, and found that perceptions of fairness, accountability, and transparency, had positive effects on perceived usefulness and convenience, mediated by trust in the system. Choung et al. (2023), also applying TAM, demonstrated that the ethical principle of trust plays a significant role in the acceptance of AI systems. This collection of studies suggests that AI ethics in HCI is not only a normative concern, but a practical one, that shapes how users perceive, trust, and engage with AI systems. This makes ethical design of AI systems central to the focus of furthering human values in HCI.

Table 1. AI Ethics Principles (Jobin et al. 2019)

| Principle | Code |
|---|---|
| Transparency | Transparency, explainability, explicability, understandability, interpretability, communication, disclosure, showing |
| Justice & fairness | Justice, fairness, consistency, inclusion, equality, equity, (non-)bias, (non-)discrimination, diversity, plurality, accessibility, reversibility, remedy, redress, challenge, access and distribution |
| Non-maleficence | Non-maleficence, security, safety, harm, protection, precaution, prevention, integrity (bodily or mental), non-subversion |
| Responsibility | Responsibility, accountability, liability, acting with integrity |
| Privacy | Privacy, personal or private information |
| Beneficence | Benefits, beneficence, well-being, peace, social good, common good |
| Freedom & autonomy | Freedom, autonomy, consent, choice, self-determination, liberty, empowerment |
| Trust | Trust |
| Sustainability | Sustainability, environment (nature), energy, resources (energy) |
| Dignity | Dignity |
| Solidarity | Solidarity, social security, cohesion |

### *Ethics Frameworks*

As AI systems become increasingly embedded in everyday life, interest in ethical AI has grown significantly within academic and policy circles (Mittelstadt 2019). Over the past decade, organisations from the private sector, civil society, government, inter-government, and multi-stakeholder organisations, have each developed their vision of ethical AI by producing a set of principles that they believe should be adhered to (Fjeld et al. 2020). These frameworks are relevant because they reflect societal norms and regulatory priorities (Papagiannidis et al. 2025), and while the documents vary in terms of their audience, scope, and detail, consensus can be found by analysing the frameworks as a whole. Several attempts have been made to review this disparate corpus of documents (Corrêa et al. 2023; Fjeld et al. 2020; Hagendorff 2020; Jobin et al. 2019; Zeng et al. 2018). In our research, we rely on (Jobin et al. 2019). The authors used the Preferred Reporting Items for Systematic Reviews and Meta-Analyses (PRISMA) framework to identify 84 published frameworks, and then manually coded ethical principles using two independent researchers. The result was a final list 63 codes, organised into 11 principles: transparency, justice and fairness, non-maleficence, responsibility, privacy, beneficence, freedom and autonomy, trust, sustainability, dignity, and solidarity (see Table 1).

Despite the importance of identifying relevant ethical principles for AI, there has been criticism that the focus on producing framework documents has distracted from moving towards implementing, or operationalising, AI ethics. This has been characterised as a ”lack of follow through” (Zhou and Chen 2023, p. 2697), merely ”good intentions” (Hagendorff 2020, p. 100), or, worse, attempts at ”virtue-signalling” intended to delay regulatory intervention (Mittelstadt 2019, p. 501). Critics point to the fact that the documents fail to resolve tensions inherent to the principles (for instance, how to provide transparency while also providing privacy), or provide any action-guiding steps that might lead towards implementing the principles (Mittelstadt 2019; Zhou and Chen 2023). Indeed, there is little evidence to suggest the body of AI ethics framework documents have meaningfully contributed to changes in the ethical design of AI systems (Mittelstadt 2019) and there are persistent challenges to meaningful implementation of the ethical principles (Papagiannidis et al. 2025).

One way of looking at the operationalisation problem is by recognising that in recent years, AI has moved from being primarily a research pursuit to one dominated by corporations (De Sousa 2025; Owens 2024). If we accept that modern AI systems are designed and developed by organisations with commercial interests (De Sousa 2025; Owens 2024), we may consider the motivation, or lack thereof, to implement the principles put forward in voluntary AI ethics framework documents. These documents are normative guidelines on what should be considered in the design and development of AI systems, because this is what is right or good. The rationale for this is philosophically derived, from, for instance, Kantian moral duties, or Aristolelian virtue ethics (Hagendorff 2020). For businesses who, it may be argued (see M. Friedman (1970)) have a primary duty to shareholder profit, this rationale is intangible and unlikely to provide meaningful motivation. This is likely why AI ethics are seen as a nice-to-have add-on, perhaps something to think about once the primary technical challenges have already been addressed (Hagendorff 2020). This view is confirmed by surveys of AI developers, who generally understand at least some of the ethical implications of their work, but report pressure to focus on performance and innovation over practical work on implementing AI ethics (Browne et al. 2024; Griffin et al. 2024; Ibáñez and Olmeda 2022; Morley et al. 2023).

### *Principles of AI Ethics*

Our research focuses on the 11 key AI ethical principles of Jobin et al. (2019), where the 11 principles are listed in the order of prevalence in global AI ethics framework documents (see Table 1). The following is a brief summary of each ethical principle as it relates to the HCI discipline.

*Transparency*. Transparency is the most commonly referenced construct in global AI ethics frameworks (Jobin et al. 2019). This may be because of its role in society more broadly, as a means of fostering knowledge and understanding in both governments (Hood and Heald 2006) and organisations (Albu and Flyverbom 2019). In AI systems, transparency can be thought of as transparency of the algorithm (Haresamudram et al. 2023), or the clarity of understanding that a human has over the decision-making processes of an AI system (Pillai 2024). Transparency is commonly operationalised through disclosure (the making explicit that an AI system is in use), explainability (the ability for an AI's decisions and functioning to be explained), understandability (the extent to which explanations are understandable to users), and explicability (the absence of vagueness in the explanation). Explanation is considered the most pertinent form of AI transparency (Zerilli et al. 2022) and has been shown to be broadly positive in how users engage with AI systems (Schemmer et al. 2022). However, organisations may be resistant to implementing transparency, due to concerns over privacy (Fjeld et al. 2020), disclosure of proprietary information (Cheong 2024), and the difficulty implementing truly useful transparency solutions (Ananny and Crawford 2018).

*Justice & Fairness*. Justice is an important ethical concept and has been central to the development of modern legal and political philosophy (Miller 2019). Justice is concerned primarily with the fairness with which people are treated (Rawls 1971, 1999). AI systems may challenge justice by delivering unfair outcomes. They may worsen socioeconomic disparities (Capraro et al. 2024), providing greater benefits to higher income individuals (Eloundou et al. 2024; Felten et al. 2023), lead to societal harms (Michael 2024; Rinta-Kahila et al. 2024), or discriminate against demographic groups (Obermeyer et al. 2019). Operationalising justice and fairness in AI systems is complex, as measuring fairness is a difficult technical challenge (Angerschmid, Zhou, et al. 2022; Mehrabi et al. 2022) and requires developers to understand complex social issues which are typically outside their area of expertise (Ibáñez and Olmeda 2022). Further complicating the issue is that user perceptions of fairness are inherently subjective (Narayanan et al. 2024). Solving this issue is important, however, as research shows perceptions of unfairness reduce trust in AI systems, making them less likely to be used, even when the user is personally advantaged by unfair bias (Angerschmid, Theuermann, et al. 2022; Draws et al. 2021)

*Non-maleficence*. A fundamental ethical principle, non-maleficence dictates that one should avoid causing any type of harm. Although not limited to physical harms, the principle of nonmaleficence is perhaps most well-known for its application in medical ethics: first, do no harm. In AI systems, we can think about this as AI safety, and limiting the potential for AI systems to cause harm to individuals, by, for example, incorrectly classifying their need for medical care (Obermeyer et al. 2019) or preventing access to employment (Dastin 2022). AI systems may endanger physical safety more directly, too, in self-driving car failures (Chougule et al. 2024) or in the generation of poisonous food recipes (Doyle 2023). They may also lead to psychological harms by promoting eating disorders (AI and Eating Disorders 2023; Fowler 2023; McCarthy 2023), and may even promote suicide to users (Guo 2025; Payne 2024). Managing these risks requires careful consideration of the potential for harm (Chan et al. 2023), and the use of strategies such as slowing down release cycles, robustness testing, external auditing, and the use of humans-in-the-loop (Bengio et al. 2024; Falco et al. 2021; Salhab et al. 2024). However, these present as tradeoffs against speed of innovation, which creates tension in a highly competitive industry. Nevertheless, nonmaleficence is important to users. Research suggests that perception of risk in using AI systems negatively affects usage intent (Choudhury 2025).

*Responsibility*. Responsibility is the moral or legal duty to act in a certain way, and to be answerable to a legal, social, or professional authority (Lucas 2004). To be answerable to one's actions is to have accountability. Responsibility is foundational to the functioning of society as it creates the link between actions and consequences. In AI systems, responsibility is focused on the idea that those who built the system should be responsible for the way it functions (Brey and Dainow 2024), and that we should know who is responsible when a negative outcome occurs (Floridi and Cowls 2019). For instance, when the algorithmic system on the Boeing 737-MAX malfunctioned in 2019 and 2020, leading to loss of life, responsibility requires that we understand who contributed to the errors in designing and certifying the system (Herkert et al. 2020). Operationalising responsiblity in AI systems may be accomplished by tracking individuals who have interfaced with the system (Santoni De Sio and Van Den Hoven 2018), and by designing systems where mistakes are identifiable so that they can be judged and corrected (K. Martin 2019). Other mechanisms include the establishment of a duty of care, and the supporting of watchdog groups and whistle-blowers (Lechterman 2022). However, absent regulatory or legal mandates, the risk of liability for negative outcomes (Kroll 2020) creates a fundamental tension which challenges the operationalisation of responsibility. Developers of AI systems must balance this, however, as perceptions of responsibility are important to users: they are associated with perceptions of trust and usefulness, which in turn drive usage intentions (Aysolmaz et al. 2023).

*Privacy*. Privacy is an important social concept dating back to ancient Greece, emerging as a legal concept in recent centuries (Holvast 2009; Westin 1970). At its core, privacy concerns the individual's right to self-determination. It is the absence of oversight in one's private affairs that allows one to be a self-determining, free-thinking individual. As it applies to AI systems, privacy is commonly thought of as informational privacy, the desire to control or influence one's personal data (Bélanger and Crossler 2011). This relates to how the system collects data, the errors it contains, secondary uses, and the potential for improper access (Smith et al. 1996). Consumers are broadly wary of privacy issues, and are known to perform a privacy calculus to decide whether to share personal information with a potential service provider (Beke et al. 2022). Additional concerns for organisations are data protection laws, which exist in some form in most developed countries (Kiesow Cortez 2021; Lim and Oh 2025), and which apply to AI systems (Curzon et al. 2021). Yet, AI systems exist in a state of fundamental tension with privacy, as they require vast amounts of personal data for the sophisticated levels of prediction and inference that is expected (K. D. Martin and Zimmermann 2024). This risks undermining the privacy interests of users.

*Beneficence*. Beneficence refers to a normative obligation, of at least some level, to help others (Beauchamp 2019). While the extent to which this obligation can be applied to individuals has been subject to great philosophical debate, beneficence is nevertheless commonplace in our modern society: universities offer scholarships, government programs support the poor, and charities serve the needs of those in developing countries. Perhaps the greatest promise of AI is its potential to further contribute to beneficence: AI may boost global productivity, invent new medicines, and tutor students, together with solving grand challenges faced by society, such as sustainability (Chen and Zhou 2022; Floridi, Cowls, et al. 2020; Nishant et al. 2020) . However, beneficence in AI is often overlooked. Societal level discussions focus on what AI shouldn't do (i.e. be harmful) more than what it should do (i.e. be good) (Singh Chauhan 2024). At an organisational level, thoughts of developing a beneficent AI system may be akin to taking a stakeholder perspective, which famously conflicts with the shareholders' right to maximise profit (De Los Reyes 2023). Yet, consumers have expectations of beneficence from organisations (Beauchamp 2019), which can further act as a source of competitive advantage (Caldwell et al.

2014). This complicates the approach organisations may take. Further research on beneficence as a factor for users of AI systems is warranted.

*Freedom & Autonomy*. Freedom can be thought of as both negative liberty, the absence of external restrictions, and positive liberty, the presence of something like self-determination that leads to freedom (Carter 2022). Freedom is deeply intertwined with autonomy, which is thought of as one's capacity to live and make choices for oneself without influence or manipulation. The two ideals are ubiquitous in modern society: our legal, political, and professional systems are all based on the assumption that an individual's preferences are their own (Colburn 2022). In AI systems, freedom and autonomy are commonly addressed in the context of online manipulation, where AI systems have the ability to undermine user autonomy and freedom by restricting choices or guiding users in a certain direction (Möhlmann 2021; Prunkl 2024; Sadeghian and Otarkhani 2024; Yeung 2017). Practical approaches to operationalising freedom and autonomy in AI systems include asking for informed consent (Andreotta et al. 2022), and giving users some form of manual input so that they can adjust AI decision-making according to their individual preferences (König 2024). Autonomy is important to users of AI systems. Reactance theory (Miron and Brehm 2006) suggests that users respond negatively where threats to their autonomy are perceived, which has been shown to be present in interactions with AI systems (Sankaran et al. 2021). *Trust*. Trust is the "intention to accept vulnerability based upon positive expectations of the intentions or behavior of another" (Rousseau et al. 1998, p. 395). It is an attitude; a subjective estimation about whether the other, another person, an organisation, or an AI system, is trustworthy, where trustworthiness is the objective, unknown measure of whether trust is warranted (McLeod 2023). Ideally, we would only trust those who are trustworthy, and only those who are trustworthy would be trusted. This makes it important that users of AI systems are able to calibrate their level of trust in AI systems effectively. Since the greatest predictors of trust in AI are performance characteristics, especially reliability (Kaplan et al. 2023), the issue of trust in AI systems is not limited to increasing user trust (that is, the subjective assessment of trustworthiness), but actually being a trust-worthy AI system, and being able to demonstrate that to users. This can be achieved through the presentation of evidence to the user so that their subjective appraisal of trustworthiness reflects the actual system capabilities over time (Rheu et al. 2021; Zhu et al. 2022). Trust in AI systems is important: it is positively associated with perceived usefulness, ease of use, and usage intention (Choung et al. 2023).

*Sustainability*. Sustainability, or sustainable development, is "development that meets the needs of the present without compromising the ability of future generations to meet their own needs" (Brundtland 1987, p. 41). Sustainability is chiefly concerned with the natural environment, the natural systems that support life, and communities, and balancing the protection of these things against the development of the human race, economy, and society (Kates et al. 2005). Getting the balance right is challenging. While AI is expected to be broadly positive for sustainability efforts (Vinuesa et al. 2020), choices made at the individual AI system can have impacts on both social and ecological sustainability (Zechiel et al. 2024). Much of this is related to data centers that power AI systems: the use of recycled materials, supply chain monitoring, renewable energy sources, all contribute to sustainability, along with the energy efficiency of the AI system itself. Factoring sustainability into the development of an AI system adds a layer of complexity, but is essential to solving this critical issue (Nishant et al. 2020). Whether the sustainability attributes of AI systems is of importance to users, however, needs further study. The green gap, a phenomenon where individuals' awareness of the importance of sustainability is not reflected in their consumption patterns, suggests this is a complex area (Elmor et al. 2024; Sukumaran and Majhi 2024).

*Dignity*. There is no single accepted definition of dignity, but may be thought of as the inescapable and inestimable worth of being human (Debes 2023; Killmister 2022) In this way, dignity separates humans from animals and, indeed, AI or machines. Dignity is closely linked with human rights: the preamble to the UN Universal Declaration of Human Rights declares the document to be based on ”recognition of the inherent dignity and of the equal and inalienable rights of all members of the human family” (Universal Declaration of Human Rights 1948, p. 1). Much of the focus of dignity in AI systems, therefore, focuses on how AI systems may somehow undermine the special quality of being human, encroach on human rights, or fail to include human-centered values. As dignity is difficult to define, measure, or test for, there are concerns about how developers are able to operationalise dignity in the design of AI systems. To address this concern, Ruster et al. (2025) developed the Dignity Lens as a method for developers to tangibly engage with dignity as a concept, which includes recognising users foundational human dignity, as well as their dignity of choice, their experience of dignity, and dignity of work. The inclusion of dignity in design may have profound impact in acceptance of AI systems. In medical contexts, patient concern about being ”reduced to a percentage” (Binns et al. 2018, p. 7) results in a sense of indignity at being treated by AI systems (Formosa et al. 2022). Further research on the effects of dignity in AI system design is warranted.

*Solidarity*. Solidarity expresses a willingness to act with others, with whom a special bond exists on the basis of shared identity or common cause (Forst 2024). Black Lives Matter, #MeToo, and the coronavirus pandemic (”we’re all in this together”) are notable examples of solidarity in action in recent times (Sangiovanni and Viehoff 2024; Stok et al. 2021), and are illustrative of the positive role which solidarity plays in protecting minority interests and building a sense of community (Van Parijs 2024). It is one of the most fundamental values in peaceful societies, yet it is not well represented in AI ethics frameworks (Luengo-Oroz 2019). This is concerning, because AI systems, similarly to digital systems generally (Russo 2024), have the potential to disrupt the human-to-human relationality that is essential to solidarity (Rudschies 2023). Operationalising solidarity requires that developers are aware that they are part of society and communities, along with industry peers, users, and stakeholders more broadly, so that this may meaningfully influence how they develop AI systems. (Rudschies 2023). The impacts of solidarity on user perceptions of AI systems has not been extensively researched.

## Methodology

Our research focuses on understanding user preferences for each of the aforementioned ethical principles. A discrete choice experiment was selected to provide answers to the research questions of this study.

### *Discrete Choice Experiments*

To elicit participant preferences for AI ethics principles, we use a discrete choice experiment (DCE). The DCE methodology was originally developed in econometrics and marketing science and is now widely used across a variety of disciplines (J. J. Louviere et al. 2000). Based on random utility theory (Daniel McFadden 1978), the DCE assumes that decision makers have latent utility from alternative attribute combinations, and that the option that maximises their utility will be their preferred choice. By observing repeated choices it is possible to statistically infer the relative preference of each attribute. DCEs are well suited to evaluating trade-offs between competing attributes, which makes them appropriate in simulating real-world scenarios where individuals must choose between alternative produces or services with a variety of different features. This makes the results of the DCE to be more reliable than other stated preference methods, lending the results greater validity.

***Ethical framework***

In order to assess consumer preferences for ethical attributes of AI systems, it was necessary to develop a discrete list of ethical attributes that reflected AI systems. Our starting point is the voluntary frameworks developed by governments, non-government organisations, and corporations. These are appropriate for our study as such frameworks often reflect societal norms and regulatory priorities (Papagiannidis et al. 2025). Given the preponderance of AI ethics frameworks globally, there have been several attempts to review extant frameworks and organise them (Corrêa et al. 2023; Fjeld et al. 2020; Hagendorff 2020; Jobin et al. 2019; Schiff et al. 2021; Zeng et al. 2018). For the purposes of this study, we have relied on the 11 principles developed by Jobin et al. (2019), which are transparency, justice and fairness, non-maleficence, responsibility, privacy, beneficence, freedom and autonomy, trust, sustainability, dignity, and solidarity.

***Instrument design***

Designing the stimuli requires representing the 11 ethical principles summarised by Jobin et al. (2019) into a format that would be easily understood by participants. Ethical principles can be difficult to understand (Ikkatai et al. 2022), therefore we designed the stimuli to capture the essence of the principle in a simple, concise, and practical bullet-point design. This approach necessitated trade-offs, since the ethical concepts are broad and complex, and cannot be meaningfully summarised in such a small amount of space. Therefore, we aimed to both capture the essential essence of the principle as it applies to consumer AI systems, and to ensure a clear differentiation between each the 11 attributes. The process followed by the researchers included several round-table review and discussion stages in order to fine-tune the stimuli. Icons were incorporated into the design as a visual heuristic to aid decision making in the repeating portion of the experiment (see Figure 1). The researchers considered the images for cultural bias, and found no major differences in underlying cultural meaning.

We were also careful in our stimuli design to avoid giving the impression that the AI systems have agency, and therefore we refer to designers and developers, and what their intent would be for the function of the AI system. For instance, for the attribute justice and fairness, we wrote ”designed to be fair in operation”, rather than ”is fair in operation”. Our approach can be considered to be taking a design stance over an intentional stance (Dennett 1998), the latter of which risks participants perceiving agentic motivation and the anthropomorphisation of the AI system, which was not the purpose of this study.

In designing the practical aspects of the discrete choice experiment, we aimed for simplicity, recognising that participants may experience high cognitive load in assessing each choice set. Thus, each choice set featured only two AI systems. Deciding how many ethical attributes that each AI system would display is a trade-off with how many choice sets are presented, since statistical efficiency declines with a larger number of attributes, but too few attributes would require a more lengthy experiment (Vanniyasingam et al. 2016). Our final specification was four attributes per choice, with two AI systems per choice set, and a total of 11 choice sets. In displaying the individual attributes, a balanced incomplete block design (BIBD) is typically used as it ensures an optimized and balanced presentation of attributes to participants (Rink 1987). However, generating a BIBD proved infeasible due to the combinatorial properties of our design. Therefore, we opted for a fully randomised design in which each attribute had the same chance of appearing in a given choice set. A post hoc check was performed to ensure that attributes were presented to participants with approximately equal frequency.

To examine how participant preferences vary by context, we utilised three emerging AI applications: AI-assisted diagnosis and treatment (context: medical), AI-based assessment of a bank loan application (context: financial), and AI-delivered education (context: educational). Each participant was assigned a single context for the duration of the experiment. To ensure participant engagement with the context they were assigned, we utilised three techniques: first, we presented a context-specific image of the physical environment which matches the application context (i.e. doctor's office, bank office, lecture theatre) throughout the task; second, we administered a brief reflection task following the vignette, but prior to the experiment; third, we included a textual reminder of the context with every choice set. This approach ensured that participants evaluated the ethical principles in their allocated context. An example of a choice set presented to participants is shown in Figure 1.

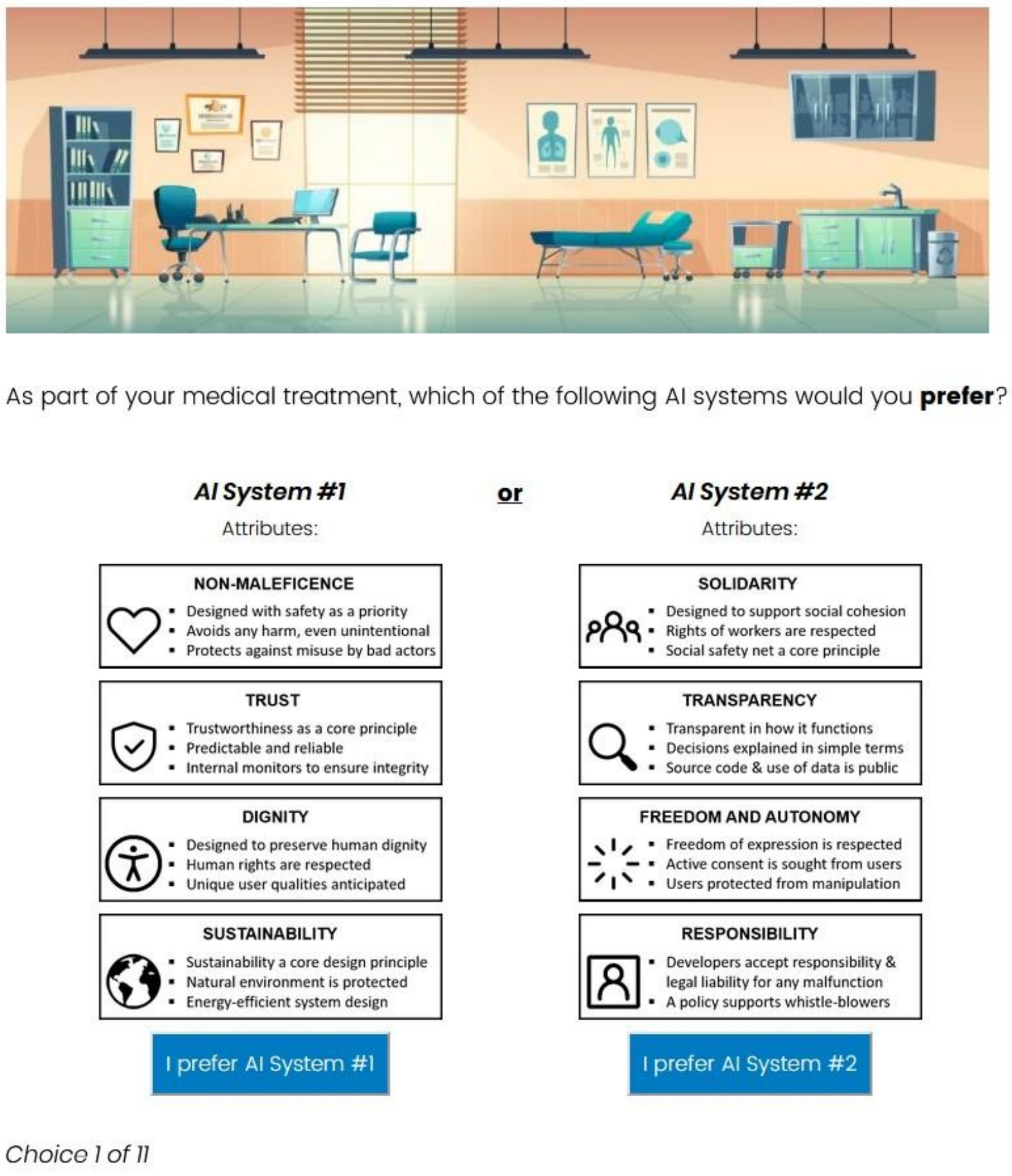

Fig. 1. An example of a choice set presented to a participant

***Translation***

The instrument was translated using the back-translation method (Brislin 1970), where the original English version is translated first to the target language (in our case, Simplified Chinese and German) by a translator, and then translated back to English by a second translator. This allows the lead investigator to compare two English-language instruments and assess for any differences that may have arisen during translation. In our study, the researchers have native fluency in at least one of the languages of the experiment, and academics external to the study were employed to provide a second translation. Where translations were found to be sub-optimal, a second round of translations were conducted, and compared once more. This was sufficient to arrive at consensus for both Simplified Chinese and German translations.

Gendered language, which can lead to participant bias in responses, was avoided in the original English. However, as German is a gendered language (Stahlberg et al. 2007), we used the suffix innen as an approach to gender neutrality, which is increasingly common in modern Germany (Waldendorf 2024). In instances where this approach was not possible, we alternated masculine and feminine forms throughout the translation for a balanced approach. Chinese, being a genderless language (Stahlberg et al. 2007), did not present the same challenges as we translated from English.

***Participants***

634 Prolific participants from the United States (US), Germany, and Australia, and 205 Credamo participants from China, completed this study for monetary compensation. Each participant was randomly assigned to one of three conditions (AI context: medical, financial, educational) in a between subjects design. Ten participants overall were excluded from participating in the study, as they either did not consent to participate, or failed an attention check at the beginning of the study. This resulted in a final sample of N = 829 ($M_{age}$ = 36.4 years; 46.4% female, 53.3% male, 0.2% preferred not to say). An overview of participants is shown in Table 2.

***Procedure***

The experiment was provided in a language which matched the participants location: participants in the US and Australia received the original English version; participants in China received a version in Simplified Chinese; those in Germany received a version in German. Upon entering the survey from their respective online platform, participants were provided with information about the experiment and asked to provide their consent to continue. They were then asked to complete an attention check question and provide basic demographic information. Next, participants were assessed on two measures relating to AI: a general measure of AI Attitudes (4-item, $\alpha$ = .864, adapted from Grassini (2023)), and a measure of AI Literacy (3-item, $\alpha$ = .851, adapted from Pinski and Benlian (2023)). A summary of the measures is shown in Table 3. The participants were then prepared for the discrete choice experiment using instructional examples, and then allocated to one of three decision contexts (AI context: medical, financial, educational). They were then presented with a short vignette relevant to the allocated context. Participants were asked to describe how they would think and feel if they were in the actual scenario presented to them. Next, the discrete choice experiment began: participants were asked to continue imagining they were in the same scenario, and they would have to choose between two AI systems with various traits. An example of such a choice is presented in Figure 1. Participants completed 11 choices, each of which was randomly generated. At the end of the 11 choices, participants were thanked for their participation in the study, and redirected to the survey platform compensation page.

## Results

***Measures***

Our two measures of AI Attitudes and AI Literacy were internally consistent: the average variance extracted (AVE) was above the .5 benchmark established by Fornell and Larcker (1981), as was the composite reliability (CR) relative to the .7 benchmark. The descriptive statistics of the measures reveal generally positive attitudes towards AI systems (M = 5.35, SD = 1.21 across the 4 items) as well as above average self-perceived literacy in using them (M = 4.75, SD = 1.31 across the 3 items). A detailed overview of the two measures is shown in Table 3.

Table 2. Sample characteristics by country

| | United States | China | Germany | Australia | Total |
|---|---|---|---|---|---|
| N | 246 | 205 | 228 | 150 | 829 |
| Gender | | | | | |
| Female | 116 | 107 | 86 | 76 | 385 |
| Male | 129 | 98 | 142 | 73 | 442 |
| Prefer not to say | 1 | 0 | 0 | 1 | 2 |
| Age | | | | | |
| 18–34 | 74 | 138 | 139 | 76 | 427 |
| 35–49 | 96 | 44 | 72 | 51 | 263 |
| 50+ | 76 | 23 | 17 | 23 | 139 |
| $\mu$ (mean) | 42.0 | 32.5 | 33.8 | 36.6 | 36.4 |
| SD | 12.9 | 10.4 | 9.9 | 12.4 | 12.1 |

***Choice Experiment***

We test the results of the choice experiment using Apollo choice modelling software (Hess and Palma 2019) which utilises the BGW algorithm (Bunch et al. 1993). Our starting point for analysis is the basic multinomial logit model (MNL). MNL is commonly used in the analysis of DCEs (J. J. Louviere et al. 2000). The probability formula can be represented as:

$$P_{ni} = \frac{exp(X'_{ni}\beta)}{\sum_{j\in J} exp(X'_{nj}\beta)} \quad (1)$$

where:
$P_{ni}$: the probability that individual $n$ chooses alternative $i$
$X_{ni}$: the vector of observed variables or attributes
$\beta$: fixed vector of coefficients

<u>Model 1.</u> Our starting point is a basic MNL model that incorporates each of our 11 attributes. The utility model, which represents how participants value the alternatives presented to them, can be represented as follows:

$$V = ASC + \sum_{k=1}^{11} \beta_k \cdot x_k + \epsilon \quad (2)$$

where:
$V$ : Systematic utility of a choice alternative
$ASC$: alternative-specific constant for a choice alternative
$\beta_k$: coefficient for attribute k
$x_k$: presence of the attribute in a choice alternative (1 for present, 0 for absent
$\epsilon$: error term, representing unobserved influences

Table 3. Measurement statistics of latent constructs

| Measure | Items (original English) | $\mu$ | SD | $\lambda$ | $\alpha$ | CR | AVE |
|---|---|---|---|---|---|---|---|
| AI Attitude | I believe that AI will improve my life | 5.18 | 1.49 | .864 | .839 | .849 | .596 |
| | I believe that AI will improve my work | 5.25 | 1.48 | .765 | | | |
| | I think I will use AI technology in the future | 5.59 | 1.63 | .792 | | | |
| | I think AI technology is positive for humanity | 5.36 | 1.26 | .589 | | | |
| AI Literacy | In general, I know the unique facets of AI and humans and their potential roles in human-AI collaboration | 4.92 | 1.37 | .851 | .895 | .896 | .743 |
| | I am knowledgeable about the steps involved in AI decision-making | 4.67 | 1.44 | .870 | | | |
| | Considering all my experience, I am relatively proficient in the field of AI | 4.67 | 1.50 | .864 | | | |

The model results are shown in Table 4. This initial model demonstrated an acceptable fit, with a Rho-squared (D. McFadden 1974) of 0.0268 (adjusted = 0.0251), log-likelihood (LL) of -6151.22, and an Akaike Information Criterion (AIC) of 12324.43, and a Bayesian Information Criterion (BIC) of 12402.73.

All but one attribute demonstrated statistically significant deviation from the baseline, indicating clear preferences for certain attributes over others. The strongest predictors, in order, were privacy ($\beta$ = 0.2258, SE = 0.0206 p < .001), justice and fairness ($\beta$ = 0.1755, SE = 0.0196, p < .001), and transparency ($\beta$ = 0.1553, SE = 0.0191, p < .001). Sustainability ($\beta$ = 0.0161, SE = 0.0185, p < .385) did not measure a statistically significant difference from the baseline of Solidarity ($\beta$ = 0.0000), indicating they are valued similarly by participants, as the least preferred of the 11 attributes.

_Model 2._ Following the estimation of our base MNL model, we investigated the possibility that unobserved preference heterogeneity or differences in scale may bias our estimates. This is important to consider in our cross-cultural study as it is possible that scale or preference heterogeneity between cultures confounds with culture-level variation in parameter estimation. To assess this possibility, we apply the scale factor estimation test from Swait and J. Louviere (1993) which distinguishes between preference heterogeneity and scale differences across respondent groups, and is computed as:

$$\chi^2 = -2\left(L_p - \sum_{g \in G} L_g\right) \quad (3)$$

where:
$L_p$: is the likelihood ratio of the MNL model specified in equation (2) using the complete pooled dataset $p$
$L_g$: is the log likelihood of the same model using the data subset $g$
$G$: US, DE, CN, AU, which represent our country-level data subsets.

The likelihood ratio statistic was significant ($\lambda$=105.48, df = 36, p < .001), indicating the presence of statistically significant scale and/or preference heterogeneity across countries. This suggests there are differences in how individuals in different countries make decisions, and this may not be captured sufficiently in the fixed-coefficient MNL model specification.

In light of this result, we considered the use of a mixed logit (MXL) model. Compared to the MNL specification, MXL relaxes assumptions such as the independence of irrelevant alternatives, and accommodates random taste variation by allowing coefficients to vary across respondents according to a normal distribution (Revelt and Train 1998). Further, it allows for scale heterogeneity (Hess and Train 2017) as identified in our likelihood ratio test. The MXL model can be represented as:

$$P_{ni} = \int \frac{exp(X'_{ni}\beta)}{\sum_{j\in J} exp(X'_{nj}\beta)} f(\beta_n \mid \theta)\, d\beta_n \qquad (4)$$

where:
$P_{ni}$: the probability that individual $n$ chooses alternative $i$
$X_{ni}$: the vector of observed variables or attributes for individual $n$
$\beta_n$: fixed vector of coefficients for individual $n$

We estimated our MXL model using the same specification as equation (2). The complete model output is shown in the appendix. While we anticipated the MXL model would meaningfully capture individual level heterogeneity, our results did not support a meaningful improvement in model fit. The log-likelihood of the MXL model (LL = -6227.01) was lower than that of the corresponding MNL model (LL = -6151.22), and thus the likelihood ratio test indicated no significant gain in explanatory power ($\chi^2(11)$ = -151.59, p = 1.000). Moreover, the MXL model produced a higher AIC (12,476.03 vs 12,324.43) and BIC (12,554.33 vs 12,402.73) compared to the MNL model. Our conclusion therefore is that the complexity of the MXL is not beneficial for our analysis. This may be due to factors such as sampling. We therefore retained the more parsimonious MNL model for all subsequent analyses.

Table 4. Overall preference levels for each AI ethical principle

| Attribute | $\beta$ | SE (robust) | p-value | sig. |
|---|---|---|---|---|
| Privacy | 0.2258 | 0.0206 | < .001 | *** |
| Justice & Fairness | 0.1755 | 0.0196 | < .001 | *** |
| Transparency | 0.1553 | 0.0191 | < .001 | *** |
| Trust | 0.1500 | 0.0186 | < .001 | *** |
| Non-maleficence | 0.1438 | 0.0201 | < .001 | *** |
| Responsibility | 0.1235 | 0.0184 | < .001 | *** |
| Dignity | 0.0773 | 0.0175 | < .001 | *** |
| Beneficence | 0.0756 | 0.0180 | < .001 | *** |
| Freedom & Autonomy | 0.0489 | 0.0185 | .008 | ** |
| Sustainability | 0.0161 | 0.0185 | .385 | |
| Solidarity | 0.0000 | n/a | n/a | n/a |

Table 5. Comparison of model fit statistics across four estimation approaches

| | Model 1 | Model 2 | Model 3 | Model 4 |
|---|---|---|---|---|
| Type | MNL | MXL | MNL | LCA |
| Parameters | 11 | 11 | 66 | 11 |
| LL | -6151.81 | -6227.01 | -6029.18 | -5859.65 |
| $Rho^2$ | 0.0268 | 0.0148 | 0.0461 | 0.0730 |
| Adj. $Rho^2$ | 0.0251 | 0.0131 | 0.0357 | 0.0655 |
| AIC | 12324.43 | 12476.03 | 12190.35 | 11813.30 |
| BIC | 12402.73 | 12554.33 | 12660.15 | 12147.85 |

*Model 3*. While the MXL model may not have provided additional exploratory power, we nevertheless sought to understand how country and context may affect choice preferences. Thus, we expanded the original MNL model with the addition of country- and context-level interactions. This expanded model can be represented as:

$$\mathrm{V} = \mathrm{ASC} + \sum_{k=1}^{11}\left[\beta_k \cdot x_k + \sum_{c\in C} \beta_k^{(c)} \cdot x_k \cdot \mathrm{Context}_c + \sum_{r\in R} \beta_k^{(r)} \cdot x_k \cdot \mathrm{Country}_r\right] + \epsilon \quad (5)$$

where:
$C$ = medical, financial, educational, which represent choice contexts
$R$ = US, DE, CN, AU, which represent participant locations

We estimated the model using Apollo. This more advanced 66-parameter MNL model reported a Rho-squared (D. McFadden 1974) of 0.0461 (adjusted = 0.0357), LL of -6029.18, AIC of 12190.35, and BIC of 12660.15. This is an improved fit over the base 11-parameter model (see Table 5 for a comparison of model fit statistics). A likelihood ratio test indicated that the latter model represented a significant improvement in model fit ($\chi^2(55) = 244.08$, $p < 0.001$). This provides statistical support for the inclusion of context- and country-level interaction effects, suggesting that they meaningfully shape preferences for the 11 AI ethics attributes.

In assessing context-level interactions, we observe that ethical preferences are broadly consistent across contexts, with several notable differences. In the medical context, non-maleficence became significantly more important relative to the aggregate baseline ($\Delta$ = +0.150, p < .001). In the finance context, non-maleficence decreased in importance ($\Delta$ = -0.093, p < .001) while justice & fairness increased in importance, becoming the most preferred attribute for this context ($\Delta$ = +0.114, p < .001). In the educational context, we find significantly lower preference for justice and fairness ($\Delta$ = -0.073, p = .006), non-maleficence ($\Delta$ = -0.057, p = .041), and transparency ($\Delta$ = -0.057, p = .032), suggesting that participants deprioritised these attributes when AI was framed as supporting learning and instruction. The model results for context-level interactions are shown in Table 6 and further presented in Figure 2.

In assessing country-level interactions, we observe that ethical preferences are broadly consistent across contexts, with several notable

Fig. 2. Attribute Preference by Context

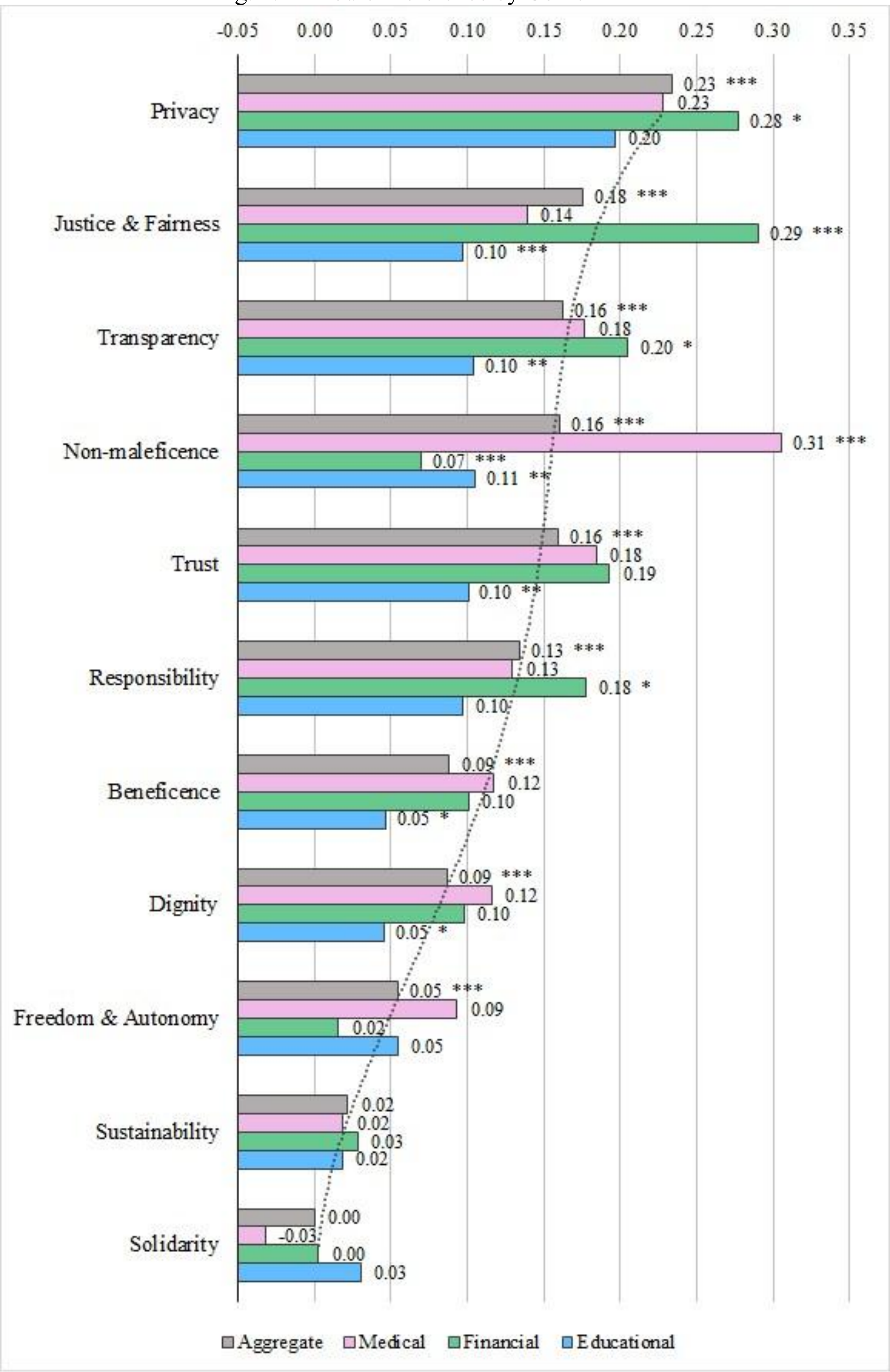

Table 6. Output from model 2 showing preferences by country gr

| Attribute | Aggregate | | US | | | DE | | | CN | | | AU | | |
|---|---|---|---|---|---|---|---|---|---|---|---|---|---|---|
| | $\beta_{aggr}$ | SE | $\Delta$ | $\beta_{US}$ | SE | $\Delta$ | $\beta_{DE}$ | SE | $\Delta$ | $\beta_{CN}$ | SE | $\Delta$ | $\beta_{AU}$ | SE |
| Privacy | 0.238*** | 0.021 | 0.027 | 0.265 | 0.054 | -0.008 | 0.229 | 0.035 | -0.153*** | 0.076 | 0.033 | 0.135* | 0.372 | 0.057 |
| Justice† | 0.182*** | 0.020 | 0.051 | 0.233 | 0.048 | 0.042 | 0.223 | 0.034 | -0.083* | 0.141 | 0.033 | -0.010 | 0.172 | 0.048 |
| Non-mal.† | 0.165*** | 0.021 | -0.071 | 0.094 | 0.050 | 0.031 | 0.196 | 0.035 | -0.157*** | 0.039 | 0.032 | 0.196*** | 0.361 | 0.055 |
| Transp.† | 0.163*** | 0.020 | -0.015 | 0.148 | 0.027 | 0.023 | 0.186 | 0.033 | -0.061 | 0.126 | 0.036 | 0.053 | 0.216 | 0.036 |
| Trust | 0.161*** | 0.020 | 0.104* | 0.265 | 0.049 | -0.002 | 0.159 | 0.031 | -0.080* | 0.079 | 0.032 | -0.022 | 0.139 | 0.054 |
| Respons.† | 0.137*** | 0.019 | -0.010 | 0.128 | 0.047 | 0.024 | 0.161 | 0.032 | -0.073* | 0.088 | 0.033 | 0.059 | 0.197 | 0.051 |
| Benef.† | 0.090*** | 0.019 | -0.003 | 0.087 | 0.046 | -0.031 | 0.059 | 0.031 | -0.043 | 0.016 | 0.032 | 0.077 | 0.167 | 0.051 |
| Dignity | 0.085*** | 0.019 | -0.006 | 0.078 | 0.030 | -0.011 | 0.074 | 0.030 | -0.033 | 0.041 | 0.030 | 0.051 | 0.135 | 0.038 |
| Freedom† | 0.055** | 0.019 | 0.067 | 0.123 | 0.047 | -0.033 | 0.023 | 0.032 | -0.075* | -0.053 | 0.033 | 0.041 | 0.096 | 0.049 |
| Sustain.† | 0.023 | 0.019 | -0.006 | 0.018 | 0.048 | -0.030 | -0.007 | 0.032 | 0.021 | 0.014 | 0.032 | 0.014 | 0.038 | 0.051 |
| Solidar.† | 0.000 | n/a | 0.028 | 0.028 | 0.083 | -0.011 | -0.011 | 0.037 | 0.020 | 0.010 | 0.041 | -0.038 | -0.038 | 0.088 |

*Note.* Asterisks denote significance levels (*p < .05, **p < .01, ***p < .001).
† indicates attribute labels which have been truncated for space.

differences. Most notable was respondents in China, who significantly deprioritised privacy (Δ = -0.153, p < .001) and non-maleficence (Δ = -0.157, p < .001) relative to the aggregate baseline. Participants in China also showed significantly lower preferences for justice and fairness (Δ = -0.083, p = .012), trust (Δ = -0.080, p = .013), and responsibility (Δ = -0.073, p = .024), suggesting a comparatively diminished emphasis on these ethical principles compared to the aggregate. Respondents in Australia, however, displayed preferences to the contrary, assigned significantly greater weight to privacy (Δ = +0.135, p = .018) and nonmaleficence (Δ = +0.196, p < .001). Respondents in the US responded in a manner closer to the aggregate baseline, only demonstrating a significantly different preference for trust (Δ = +0.104, p = .032). Participants in Germany were found to be the most similar to the aggregate, with no significant attribute-level differences. Across all four countries, differences in preference weights for Sustainability and Solidarity were consistently non-significant. indicating a low level of preference.

_Model 4_. While our expanded MNL model provides valuable insight into how context and country moderate preferences for ethical attributes of AI systems, it may still not adequately account for individual heterogeneity. To address this, we implemented a Latent Class Analysis (LCA) to segment respondents into a discrete number of unobserved latent classes, each of which are characterised by their own preferences for AI ethical attributes. This approach allows us to explore unobserved heterogeneity as an alternative to the MXL model, which did not yield reliable results in our analysis. Our primary motivation for adopting LCA was to test whether individual-level characteristics, such as age, gender, AI attitudes, and AI literacy, are predictive of class membership more so than context or culture.

We followed the recommended procedure in Apollo and used distinct coefficient starting values to ensure an identified global maximum. The base MNL model of equation (2) was used to estimate preference classes. In a procedure matching Burke et al. (2010), we iterated model estimations with the aim of identifying the number of classes which minimised $BIC = -2L + m \cdot log(n)$ where $L$ is the log-likelihood of the model with $m$ parameters and $n$ sample size. BIC is appropriate for assessing latent class model fit and is superior to AIC for this purpose (Yang 2006). The BIC of the four-class model was found to be 12147.85, lower that the three-class model and the five-class model, and thus was selected as the preferred number of classes. Adding support to this decision is that four classes is typically appropriate for our sample size of N = 829 (Nylund et al. 2007). The 4-class LCA model was found to be to be superior fitting when compared to the previous MNL models. In addition to the aforementioned BIC, the 4-class LCA model reported a Rho-squared (D. McFadden 1974) of 0.0730 (adjusted = 0.0655), LL of -5859.65, and an AIC of 11813.30. The LCA is therefore the best fitting model on all measures (a summary of models is shown in Table 5).

Following class estimation, individual-level posterior probabilities of class membership were calculated, which represent the likelihood of a respondent belonging to each of the four latent classes estimated, based on their observed choice behaviour. Using these posterior probabilities as frequency weights, we then estimated a multinomial logistic regression model using individual level decision context, demographic, sociographic, and psychographic variables. This allowed the interpretation of the regression coefficients as predictors of class membership. The results of the latent class analysis and multinomial regression are presented in Table 6, along with descriptive labels assigned by the researchers.

## Discussion

This study set out to empirically evaluate user preferences for each of the AI ethical principles represented in global framework documents. In doing so, we aimed to contribute to the need to a clear pathway to impact (Mittelstadt 2019) and contribute to the ongoing operationalisation of

Table 7. Latent class model estimates across four participant segments

| | Class 1 | Class 2 | Class 3 | Class 4 |
|---|---|---|---|---|
| Class label | Privacy-centric | Balanced moderates | Ethically disengaged | Safety conscious |
| Class share | 13% | 32% | 47% | 8% |
| *Attribute Preferences* | | | | |
| Transparency | 0.41*** | 0.42*** | 0.01 | 0.30*** |
| Justice & Fairness | 0.21* | 0.59*** | 0.03 | 0.54** |
| Non-maleficence | 0.13 | 0.18*** | 0.02 | 2.08*** |
| Responsibility | 0.27*** | 0.14* | 0.10** | 0.71*** |
| Privacy | 1.57*** | 0.24*** | 0.01 | 0.67*** |
| Beneficence | 0.06 | 0.18*** | 0.01 | 0.48*** |
| Freedom & Autonomy | 0.18* | 0.13* | 0.02 | 0.36** |
| Trust | 0.53*** | 0.31*** | 0.04 | 0.64*** |
| Sustainability | 0.30*** | 0.14** | 0.06 | 0.34** |
| Dignity | 0.25** | 0.17*** | 0.02 | 0.51*** |
| Solidarity | 0.00 | 0.00 | 0.00 | 0.00 |
| *Membership Predictors* | | | | |
| Intercept | 0.149*** | 0.283*** | 0.472*** | 0.096*** |
| Context | | | | |
| Medical | -0.032* | -0.031* | -0.012 | 0.074*** |
| Financial | 0.007 | 0.078*** | -0.026 | -0.059*** |
| Educational | 0.025 | -0.047** | 0.038* | -0.015 |
| Location | | | | |
| USA | 0.015 | 0.010 | 0.024 | -0.049*** |
| Germany | 0.003 | 0.019 | -0.068** | 0.047*** |
| China | -0.074*** | -0.019 | 0.164*** | -0.071*** |
| Australia | 0.056** | -0.011 | -0.119*** | 0.073*** |
| Demographics | | | | |
| Age | -0.008 | 0.018 | -0.015 | 0.005 |
| Female | 0.011 | -0.014 | -0.006 | 0.009 |
| Psychographics | | | | |
| AI Attitude | -0.008 | 0.009 | -0.004 | 0.002 |
| AI Literacy | 0.006 | -0.021 | 0.019 | -0.004 |

Note. * indicates significance at the $p < .05$ level; ** $p < .01$; *** $p < .001$. Attribute Preferences: significance is calculated relative to the arbitrary baseline represented by the attribute solidarity. Membership predictors: as the result of a linear regression, significance is interpreted more traditionally as an indicator of confidence in the predictor values.

AI ethics. To achieve this, we sought to understand the preferences of AI ethical preferences amongst end users of AI systems, and whether they varied meaningfully between culture, context, and individual traits.

Using a discrete choice experiment (DCE), we have shown compelling evidence that some principles are relatively more important than others. Privacy, justice & fairness, and transparency have emerged most important ethical concerns for users of AI systems. Sustainability and Solidarity, meanwhile, were deprioritised by participants in our study. This lends important empirical evidence in support of RQ1, where we sought to understand whether some principles would be more favoured than others. Overall, these findings support the notion that some ethical principles are more important to users than others, and justify uneven focus in operationalision priority.

These preferences however were not uniform across context or culture, as anticipated by RQ2 and RQ3. We found significant deviation from the aggregate preference baseline for one or more ethical principle in each of the three contexts we assessed. In the financial context, justice and fairness was the most important for respondents, suggesting concern that they would be treated without bias in the application process, while in the medical context, the primacy of non-maleficence suggested concern over the safety of AI-powered medical systems. This suggests that the context of the AI system operates meaningfully shapes which principles participants valued most, and would be most responsive to as system attributes, and provides empirical support to the context-specific differences suggested by Zhou and Chen (2023).

We also found significant deviation from the aggregate when assessing country-level preferences. This was particularly notable in China, where privacy and non-maleficence were significantly muted in overall preference when compared to the three Western countries represented in our experiment. Instead, Chinese participants had an overall preference for justice & fairness and transparency. This suggests the cultural environment in which the AI system operates meaningfully shapes which principles participants value most. This supports the findings of Corrêa et al. (2023), who, in analysing global approaches to AI ethics, noted differences in how principles where categorised across countries. The need to move beyond a globally hegemonic approach to AI ethics has been noted in prior literature (for instance, (Hongladarom and Bandasak 2024)). Our research contributes to this discussion by quantifying preferences at the individual user level across four countries, providing empirical evidence of cultural heterogeneity in AI ethics.

We did not find a significant effect of either general AI attitudes or AI literacy (RQ4 and RQ5), indicating these factors did not have a meaningful effect on attribute preference.

The robustness of the overall findings were shown through a latent class analysis (LCA), which accounted for unobserved individual-level heterogeneity by organising participants into classes according only to their attribute preferences. The LCA revealed four main segments of users, which we labelled according to the relative preferences of each class. The privacy-centric class accounted for 13% of participants, and displayed a clear and singular focus on privacy as the most preferred ethical attribute. Australian participants were most likely to be part of this group, and those not in the medical context. The safety-conscious class, 8% of participants, has a focus on non-maleficence. Participants in this class were most likely to be in the medical context, or from the German or Australian cohort generally. The balanced moderates class, which was 32% of participants, found all attributes other than solidarity to be important, but without the same intensity as shown in the two preceding classes. Participants in this group were more likely to be in the financial context. Finally, the largest class, ethically disengaged, which accounted for 47% of participants. Preferences of participants in this group were overall muted, but with a special focus on responsibility. This may suggest that for this segment of the population, there is an implicit trust in institutional oversight, and that they engaging deeply with ethical concerns is unnecessary. Chinese participants and those in the educational context are most likely to be found in this group.

The LCA confirmed the general findings out our study, that there is no single approach to AI ethics that will be universally accepted amongst a diverse user base. However, it also revealed that while there are small groups of ethically passionate users, most users are either not concerned with the individual makeup of ethical principles as long as the system is ethical overall (the balanced moderates), or they are not interested in ethical principles at all as long as there is responsibility embedded into the system design (the ethically disengaged). This later group is the largest, and accounts for almost half of our participants. This suggests that an alternative approach to AI ethics operationalisation, that would be accepted by users, may be in the form of a special focus on responsibility. Such an approach would focus on oversight mechanisms to ensure accountability: regulation, the establishment of a professional body for AI developers, certification, the development of codes of conduct, and the ascribing of formal liability (Cheong 2024; Dignum 2020; Kroll 2020; Lechterman 2022). Our research contributes to this discussion by showing that focusing on responsibility as an AI ethics operationalisation strategy may be acceptable to a large percentage of users of AI systems.

### *Implications*

These findings offer important insights for HCI or Human-AI Interaction (HAII) researchers and practitioners. First, they highlight the potential benefits of integrating user preferences for ethical principles into the design of AI systems. By identifying which principles matter the most to users in a given application and cultural context, designers and developers can better prioritise the features that will resonate with the actual demands of their users. For instance, in health systems intended for Western users, strengthening mechanisms for non-maleficence and privacy, and communicating that to users, may yield greater intention to use and adopt the AI system. This provides a method of effectively improving outcomes using ethical design.

Second, this research helps make the case for ethical design factors. This knowledge of user preferences for ethical principles allows designers and developers to more confidently make trade-offs, both with competing ethical priorities, as well as overall business objectives. Continuing the above health context example, sustainability and solidarity, which are lower priority to users, can appropriately be de-prioritised by developers so that they are able to focus on achieving more meaningful outcomes on the key principles of privacy and non-maleficence. At the business level, since these attributes are so overwhelmingly preferred, their inclusion may justify a greater organisational focus, which influences resource investments into ethical design, perhaps in lieu of additional features. The quantified, empirical evidence of our research supports the methods by which commercial organisations make decisions about organisational priorities. This is one of the key implications of our research.

Third, these results provide actionable motivation for ethical AI design in practice, by showing that users see ethical principles not as abstract ideas, but as features which they would like to see embedded in AI systems. This moves the discussion of AI operationalisation from one of abstract ideals or regulatory compliance, to one of feature design. AI ethicists and those supportive of ethical AI systems can point to user preferences as a pathway to system acceptance and adoption, which may be more effective that existing arguments based on normative standards justified by moral idealism. This provides a clear pathway to impact which has been missing in the AI ethics operationalisation discussion to date.

Fourth, and finally, our research confirms the need for culturally responsive and domain-sensitive design in AI systems. There is no single universal approach to AI ethics, owing to the diversity of humanity and application contexts of AI technology. However, there is a risk that, in the development of AI systems, particularly those capable of rapid scalability across multiple languages and domains, the importance of this may be overlooked. Our research makes clear that there is no one-size-fits-all approach to implementing AI ethics principles. HCI practitioners and AI designers must consider how localised user priorities will shape the use and acceptance of the AI system they are developing.

***Limitations***

This study, while comprehensive in scope, is subject to several limitations. First, the discrete choice experiment assumes the independence of irrelevant alternatives, which holds that the choice between two given AI systems would not be influenced by a third option being known. In practice, this may not be realistic. The design also assumes independence of attributes, but for which we know to be conceptually and practically intertwined (for instance, a breach of privacy may impact non-maleficence). Our model does not account for such inter-dependencies, or measure if the participants perceive there to be such connections.

Second, while the DCE is a robust stated choice methodology, it suffers from the same drawbacks of many experimental designs in that it does not capture actual behaviours. This may be particularly pertinent when assessing the results of our study, which reveals privacy to be the most important ethical attribute to participants. Yet, the privacy paradox shows that people who value privacy will nevertheless still use digital platforms that put their privacy at risk (Gerber et al. 2018). This is just one example of the phenomenon known as the intention-behaviour gap. Ultimately, further research is needed to understand how this applies to behavioural adoption and use of AI systems.

Thirdly, our design did not utilise negatively framed attributes (i.e. ethical breaches), which may have provided a greater richness to the discussion. However, our focus for this study was from the perspective of designers and developers who consider which features to incorporate into AI systems. In keeping with how system features are designed and presented, our study focused on positively framing each principle. A broader perspective and avenue for future research may be to consider the risk of unethicality, and how this affects user intentions and behaviours.

Finally, while our study captures perceptions of ethical principles, it does not consider the certainty with which perceptions are formed, which in actual applications may be an important factor in how users assess such AI systems. For instance, if an AI system is marketed as 'privacy focused', the user may yet consider whether merely this is marketing-speak, and question whether the developers have actually followed a privacy-by-design development path, and then further there is still the risk that the system is not working as intended, or has a security flaw that will lead to an data leak in the future. Should the developers of AI systems simply focus on the marketing of AI ethics, we risk not making progress on the operationalisation problem. This issue is seen in other areas of ethical concern, such as in sustainability, where greenwashing describes an organisation which communicates positively about sustainability while acting in the opposite fashion (De Freitas Netto et al. 2020). This risk may cloud potential positive impacts of a user-first approach to AI ethics operationalisation.

Despite these limitations, our findings offer a robust empirical foundation for understanding user preferences for AI ethics, and a user-centered lens for guiding ethical AI design.

## Conclusion and Future Work

Overall, our research supports the importance of cultural and context-specific approach to AI ethics, and suggest that globally standardised ethical frameworks may fail to resonate uniformly across user populations. Our research provides a clear contribution to literature by providing empirical evidence of heterogeneity in user preferences. Our findings support ongoing discussions in HCI and HAII literature, and provide practical guidance to practitioners on how to approach the challenges associated with AI ethics operationalisation.

Future work may wish to focus on individual factors associated with AI ethical principles. Extant research suggests individual factors are influential in how users build trust, form intent to use, and gain satisfaction from AI systems (for instance, see Gursoy et al. (2019) and Kaplan et al. (2023)). However, our research shows that two key individual traits (AI attitude and AI literacy) were not significant factors in either attribute preference or class membership. While it is possible that the way in which individual's develop preferences for ethical principles is distinct, it may also be the case that this effect is minor when compared to the primary effect of context and culture, and that individual traits were unobservable in our models. Further research on this topic would be valuable.

Future research may also extend the examination of user preferences for embedded AI ethics to additional countries and application contexts. While our findings are robust and point to clear preferences across countries and contexts, the generalisability of these findings remains bounded by the scope of our study. Replicating this work in additional contexts and countries could strengthen the evidence base, while potentially uncovering novel preferences that are of value to both researchers and practitioners.

## Acknowledgments

The authors would like to thank Wenzel Mehnert, Axuan Sun, Christine Eckert, and Zhibo Jin for their assistance in translating components of the study. This research has been financially supported by the University of Technology Sydney (UTS) Key Technology Partnership (KTP) Seed Funding Scheme 2024, and by an Australian Government Research Training Program (RTP) Scholarship.

**Appendix:** Table 8. The 11 attributes and translations

| English | Simplified Chinese | German |
|---|---|---|
| TRANSPARENCY<br>• Transparent in how it functions<br>• Decisions explained in simple terms<br>• Source code & use of data is public | 透明度<br>• 它如何运作是透明的<br>• 用简单的术语解释决策<br>• 源代码和数据使用是公开的 | TRANSPARENZ<br>• Transparent in der Funktionsweise<br>• Entscheidungen in einfachen Worten erklärt<br>• Quellcode & Nutzung der Daten sind öffentlich |
| JUSTICE AND FAIRNESS<br>• Designed to be fair in operation<br>• Prevention of bias or discrimination<br>• It is possible to appeal AI decisions | 正义与公平<br>• 旨在公平运作的设计<br>• 预防偏见或歧视<br>• 可以对 AI 的决策提出上诉 | RECHT UND FAIRNESS<br>• Entwickelt, für eine faire Funktionsweise<br>• Verhinderung von Vorurteilen und Diskriminierung<br>• KI-Entscheidungen können angefochten werden |
| NON-MALEFICENCE<br>• Designed with safety as a priority<br>• Avoids any harm, even unintentional<br>• Protects against misuse by bad actors | 无恶意<br>• 以安全为优先的设计<br>• 避免即使是无意的伤害<br>• 防止不良行为者滥用 | KEIN SCHADEN<br>• Entwickelt mit Sicherheit als Priorität<br>• Vermeidet jeglichen Schaden, selbst unabsichtigen<br>• Schutz vor Missbrauch durch böswillige Akteure |
| RESPONSIBILITY<br>• Developers accept responsibility & legal liability for any malfunction<br>• A policy supports whistle-blowers | 责任<br>• 开发者对任何故障承担责任和法律责任<br>• 支持举报人的政策 | VERANTWORTLICHKEIT<br>• Die Entwickler:innen übernehmen die Verantwortung und rechtliche Haftung für jegliche Fehlfunktion<br>• Eine Richtlinie unterstützt Whistleblower |
| PRIVACY<br>• User privacy is respected<br>• Privacy rights a core design principle<br>• User data is secure & protected | 隐私<br>• 尊重用户隐私<br>• 隐私权是核心设计原则<br>• 用户数据安全且受到保护 | PRIVATSPHÄRE<br>• Die Privatsphäre der Nutzer:innen wird respektiert<br>• Recht auf Privatsphäre als zentrales Gestaltungsprinzip<br>• Benutzerdaten sind sicher und geschützt |
| BENEFICENCE<br>• Designed to "do good"for users<br>• Enhances well-being and happiness<br>• Fosters socio-economic opportunity | 仁爱<br>• 为用户" 行善" 的设计<br>• 增强福祉和幸福感<br>• 促进社会经济发展 | GEMEINWOHL<br>• Entwickelt, um „Gutes"für die Nutzer:innen zu tun<br>• Verbessert das Wohlbefinden und die Zufriedenheit<br>• Fördert die sozioökonomischen Möglichkeiten |
| FREEDOM AND AUTONOMY<br>• Freedom of expression is respected<br>• Active consent is sought from users<br>• Users protected from manipulation | 自由与自治<br>• 言论自由受到尊重<br>• 征求用户的主动同意<br>• 保护用户免受操纵 | FREIHEIT UND AUTONOMIE<br>• Das Recht auf freie Meinungsäußerung wird geachtet<br>• Die aktive Zustimmung der Nutzer:innen wird eingeholt<br>• Nutzer:innen sind vor Manipulation geschützt |
| TRUST<br>• Trustworthiness as a core principle<br>• Predictable and reliable<br>• Internal monitors to ensure integrity | 信任<br>• 值得信赖作为核心原则<br>• 可预测且可靠<br>• 内部监测以确保完整性 | VERTRAUEN<br>• Vertrauenswürdigkeit als Grundprinzip<br>• Vorhersehbar und zuverlässig<br>• Interne Kontrollen, um Integrität zu gewährleisten |
| SUSTAINABILITY<br>• Sustainability a core design principle<br>• Natural environment is protected<br>• Energy-efficient system design | 可持续性<br>• 可持续性是核心设计原则<br>• 自然环境得到保护<br>• 高效能耗的系统设计 | NACHHALTIGKEIT<br>• Nachhaltigkeit als zentrales Gestaltungsprinzip<br>• Die natürliche Umwelt wird geschützt<br>• Energieeffizientes Systemdesign |
| DIGNITY<br>• Designed to preserve human dignity<br>• Human rights are respected<br>• Unique user qualities anticipated | 尊严<br>• 维护人类尊严的设计<br>• 人权受到尊重<br>• 预期的独特用户品质 | WÜRDE<br>• Gestaltet zur Wahrung der Menschenwürde<br>• Die Menschenrechte werden berücksichtigt<br>• Einzigartigkeit jedes Nutzers wird berücksichtigt |
| SOLIDARITY<br>• Designed to support social cohesion<br>• Rights of workers are respected<br>• Social safety net a core principle | 团结<br>• 支持社会凝聚力的设计<br>• 工人的权利受到尊重<br>• 社会安全是核心原则 | SOLIDARITÄT<br>• Fördert den sozialen Zusammenhalt<br>• Die Rechte der Arbeitnehmer:innen wrden gewahrt<br>• Soziale Sicherheit als Kernprinzip |